\documentclass[11pt, letterpaper]{article}
\usepackage[utf8]{inputenc}
\usepackage[margin=1.5cm]{geometry}
\usepackage{titlesec}
\usepackage{tabu}
\usepackage{enumitem}
\usepackage{amssymb}
\usepackage{xcolor}

\usepackage{graphicx}
\usepackage{lscape}
\usepackage{adjustbox}
\usepackage{geometry}
\usepackage{afterpage}

\newlist{selectlist}{itemize}{2}
\setlist[selectlist]{label=$\square$,leftmargin=*,noitemsep,topsep=0pt}

\usepackage{lmodern}

\usepackage{hyperref}
\hypersetup{
    colorlinks=true,
    linkcolor=blue,
    filecolor=magenta,      
    urlcolor=blue,
}
 
\titleformat{\section}[block]{\hspace{1em}\bfseries}{\thesection.}{0.5em}{} 
\titleformat{\subsection}[block]{\hspace{1em}}{\thesubsection}{0.5em}{}

\begin{document}
\begin{flushleft}

\setlength{\parindent}{0pt}
\setlength{\parskip}{10pt}

\textbf{Article title}\\ \textit{A simple method for deriving the birdcage coil magnetic field with experimental validation at 4 T, 7 T and 15.2 T}

\textbf{Authors}\\ A. Villareal\textsuperscript{1}, J. Lazovic\textsuperscript{2}, S. E. Solis-Najera\textsuperscript{1}, R. Martin\textsuperscript{1}, R. Ruiz\textsuperscript{1}, L. Medina\textsuperscript{1}, A. O. Rodriguez\textsuperscript{3}(*)

\textbf{Affiliations}\\ \textsuperscript{1} Departamento de Fisica, Facultad de Ciencias, UNAM, Mexico City 04510\\ \textsuperscript{2} Department of Physical Intelligence, Max Planck Institute for Intelligent Systems, Stuttgart 70569, Germany\\ \textsuperscript{3} Department of Electrical Engineering, UAM Iztapalapa, Mexico City 09340. Mexico.

\textbf{Corresponding author’s email address and Twitter handle}\\ correo@correo.com

\textbf{Abstract}\\ Magnetic resonance imaging and spectroscopy rely on the magnetic fields generated by radiofrequency volume coils to acquire high-quality data. Consequently, a comprehensive understanding of electromagnetic field behavior in RF volume coils is essential for optimizing imaging techniques and designing advanced coils. This study introduces a theoretical model for the magnetic field generated by a birdcage coil, based on a spherical geometry approach. To validate the proposed model, phantom images were acquired at different resonant frequencies, and the magnetic field produced by the RF coil was compared with experimental data. The results demonstrate the accuracy and effectiveness of the theoretical model, offering valuable insights into the behavior of electromagnetic fields in RF coils. This study provides a promising framework for
further analysis and development of RF coil design, with significant implications for advancing both MRI and spectroscopy technologies.

\textbf{Keywords}\\ magnetic resonance imaging, RF volume coil, birdcage resonator, theoretical magnetic field, ultra high field, preclinical hardware.

\newpage
\textbf{Specifications table}\\
\vskip 0.2cm
\tabulinesep=1ex
\begin{tabu} to \linewidth {|X|X[3,l]|}
\hline  \textbf{Hardware name} & birdcage coil for high and ultra high field at 4T, 7 T and 15.2 T
  \\
  \hline \textbf{Subject area} & Magnetic Resonance Imaging, Biomedical Engineering
  \\
  \hline \textbf{Hardware type} & Imaging devices
  \\ 
\hline \textbf{Closest commercial analog} & Commericial birdcage coils
  \\
\hline \textbf{Open source license} &   Creative Commons Attribution-ShareAlike 4.0 International License (CC BY-SA 4.0)
  \\
\hline \textbf{Cost of hardware} & 250 USD
  \\
\hline \textbf{Source file repository} & Available in the article
\\\hline
\end{tabu}
\end{flushleft}
\newpage
\section{Hardware in context}
Different types of RF coils are employed in MRI systems, depending on the specific requirements of the scan, such as coils designed for particular body regions or specialized imaging techniques. RF coils are essential components of the MRI process, as they enable the generation of high-resolution images that are critical for diagnosing and evaluating a wide range of medical conditions [1]. Key factors such as coil sensitivity and uniformity play a crucial role in the performance of these devices, particularly when designing coils for diverse applications.
Among the various RF coil designs, birdcage (BC) coils are particularly favored in MRI due to their ability to provide a high signal-to-noise ratio and excellent homogeneity of the RF magnetic field [2].
These characteristics ensure a large, uniform field of view, making BC coils ideal for imaging large anatomical regions or for applications requiring high spatial resolution. Despite advances in coil technology over the years, BC coils remain a primary choice for both preclinical and clinical MRI applications due to their reliability, versatility, and superior imaging performance [3-5]. 

In this paper, we present a classic theoretical approach to derive an expression for the magnetic field B\textsubscript{1} generated by a BC coil. We intend to use the formula for: a) assessing RF coil performance to acquire images of optimal quality, ensuring high signal-to-noise ratios (SNR) and accurate anatomical representation, and b) guiding the development of BC coils tailored to meet the specific needs of various applications.

Our method is based on the expansion of first- and second-order spherical Bessel functions, which provides an efficient means of characterizing the electromagnetic field within the BC coil. This approach simplifies the mathematical complexity typically associated with the analysis of RF coils, making it easier to predict the spatial distribution of the magnetic field. To validate the accuracy and effectiveness of the derived expression, we conducted a series of experiments involving the acquisition of phantom images using both in-house and vendor-provided BC coils across multiple resonant frequencies. We then compared the experimental data obtained from these measurements with the theoretical results derived from our magnetic field expression.

\section{Theory}

Bidinosti et al. [6] developed an expression for the magnetic field, $\mathrm{B}_{1}$ produced by an RF volume coil within a spherical volume, extending the theoretical framework established by London [?]. Their formulation accommodates a sphere with arbitrary permeability $\mu$, permittivity $\varepsilon$, and electrical conductivity $\sigma$.

We only considered the $\mathrm{B}_{1}$ inside the sphere because this the important concern in MRI experiments. The sphere has radius \emph{a} and is located at the origin in a uniform applied RF field, $\mathbf{B}_1(t) = B_1 \mathrm{e}^{-i\omega t} \hat{\mathbf{e}}_z$, the magnetic field inside the sphere is

\[
\mathrm{B}_{1}\left(r,\theta\right)=\frac{2J_{\psi}}{3i\omega\sigma}\frac{kj_{0}(ka)}{j_{1}(kr)\sin\theta}\quad(1)
\]

The complete derivation of eq. (1) is provided in Appendix A. 
\section{Hardware description}
The study utilized both vendor-provided and in-house BC coils, ensuring a comprehensive evaluation of performance and design variations between commercially available and custom-built models.

a) High-Pass Birdcage Coil at 300 MHz:

A high-pass birdcage coil with a diameter of 40 mm and a length of 64 mm was designed with four rungs to achieve a low specific absorption rate (SAR), as detailed in [8]. The coil's diameter-to-length ratio of 0.625 was optimized for enhanced field homogeneity and minimized signal-to-noise ratio (SNR) degradation. Operating in quadrature mode and in transceiver configuration, the coil demonstrated efficient performance, making it well-suited for the intended application.

b) Low-Pass Birdcage Coil at 300 MHz:

This low-pass birdcage coil [9] was carefully constructed using high-quality copper strips for the four rungs, mounted symmetrically along the perimeter of a durable acrylic cylinder. The ends were secured with precision-machined circular end rings. Resonance was achieved through fixed-value, non-magnetic chip capacitors (American Technical Ceramics, series ATC 100 B, 8.8 pF and 7.8 pF), which were soldered onto the end rings. These capacitors enabled stable resonant modes for both birdcage and rectangular slot coil configurations. A well-designed tuning and matching network was integrated, with BNC connectors for seamless signal transmission. For precise 50 $\Omega$ matching and fine-tuning, two non-magnetic trimmers (Voltronics Corp: 1--33 pF, NMAJ30 0736) were added, one for each channel. The quadrature-driven transceiver coil delivered improved signal uniformity, reduced artifacts, and enhanced performance, making it ideal for high-resolution imaging and other advanced MR applications.

c) BC Coil for 300 g Rats at 300 MHz:

The high-pass birdcage coil [10] for use with rats weighing approximately 300 g was designed with 16 rungs to achieve optimal $\mathrm{B}_{1}$ field uniformity, as recommended by Doty et al. [11]. Copper strips were evenly distributed around the perimeter of an acrylic cylinder, with the ends connected by two circular copper end rings. Non-magnetic chip capacitors (American Technical Ceramics, series ATC 100 B, 18 pF) were soldered to the end rings to facilitate resonance. A comprehensive tuning and matching network was implemented, with BNC connectors at both ends. Fine-tuning was achieved using two non-magnetic 1--33 pF variable capacitors (Voltronics Corp, NMAJ30 0736) for each channel.
To further enhance performance, cable traps were incorporated between the BNC connectors and matching capacitors. These cable traps, constructed from coiled coaxial cable with eight turns and a 1.5 cm inner diameter, were equipped with variable capacitors to adjust resonance to the 300 MHz Larmor frequency. The quadrature-driven transceiver coil optimized signal uniformity, reducing artifacts and ensuring high-quality MR
imaging.

d) Vendor-Provided Birdcage Coil for Comparison at 300 MHz:

For comparison, a vendor-provided quadrature birdcage coil operated in transceiver mode was used [12]. This coil had similar dimensions and an equal number of rungs to the custom-designed models. It was the RF RES 300 1H 075/040 QSN TR (model no.: 1PT13161V3, serial no.: S0121, REV/VEC: 2P01.05, Bruker BioSpin MRI, GmbH, Germany) and provided performance benchmarks against the custom prototypes.

Birdcage Coil at 170 MHz:

A pass-band birdcage coil was developed for operation at 170 MHz, as shown in Fig. 2a. This coil, designed with an optimized configuration based on [13], had a length of 12 cm and an inner diameter of 18 cm. The 4-leg architecture utilized 20 mm wide copper strips for the legs, and the circular end rings had a diameter of 180 mm. For quadrature operation, two 50 $\Omega$ coaxial cables were connected to the coil. Non-magnetic capacitors (23.2 pF and 12 pF) were symmetrically distributed around the coil's circumference. Matching was achieved using two non-magnetic 30 pF variable capacitors for fine-tuning to 170.29 MHz. The 40 mm spacing between the copper strips minimized mutual inductance effects, ensuring high-performance operation. This meticulously engineered design delivered stable and efficient MR imaging, optimized for precise field homogeneity and minimal interference.

Birdcage Coil at 650 MHz:

A quadrature transceiver birdcage coil [14] was employed for 650 MHz operation at 15.2 T (Bruker Co, Ettlingen, Germany, model RES 650 1H 059/035 QSN TR), with an inner diameter of 35 mm, outer diameter of 59 mm, and a length of 30 cm. This high-frequency coil was used for advanced MR imaging, offering excellent performance in high-field applications.

Fig. 1 presents photographs showcasing all the BC coils, providing a visual comparison of their designs and construction details.

\medskip{}
\noindent\begin{minipage}[t]{1\columnwidth}%
\begin{center}
\includegraphics[scale=2]{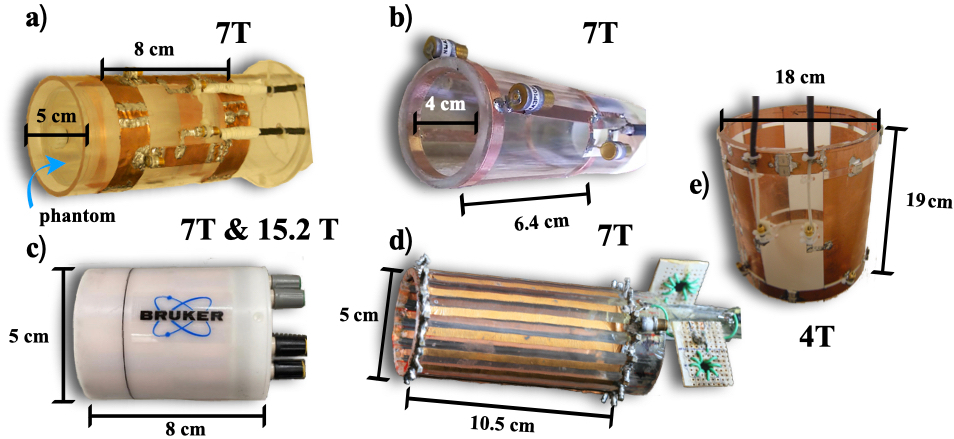}
\par\end{center}
\begin{center}
Figure 1. Photographs of the vendor-provided (c) and in-house (a, b, d, e) birdcage coils are shown, highlighting their dimensions for comparison.
\par\end{center}
\end{minipage}

\medskip{}

\section{Bill of materials summary}
\begin{center}
\begin{tabular}{|c|c|c|c|c|c|}
\hline 
{\tiny{}Designator/ name } & {\tiny{}Number} & {\tiny{}Cost/unit {[}US dollars{]} } & {\tiny{}Total cost } & {\tiny{}Source of Material } & {\tiny{}Material type }\tabularnewline
\hline 
\hline 
{\tiny{}copper sheet } &  & {\tiny{}4/1 m} &  & {\tiny{}Aceros y Metales Cuatitlan} & {\tiny{}copper}\tabularnewline
\hline 
{\tiny{}Self adhesive Copper tape 25.4 mm } &  & {\tiny{}20/1} &  & {\tiny{}3M Mexico} & {\tiny{}copper}\tabularnewline
\hline 
{\tiny{}Ceramic capacitor } &  & {\tiny{}6.5/1} &  & {\tiny{}Mouser Electronics} & {\tiny{}semiconductor}\tabularnewline
\hline 
{\tiny{}Coaxial cable (50 Ohm impedance)} &  & {\tiny{}5/1 m} &  & {\tiny{}PCdigital} & {\tiny{}silver coated metal}\tabularnewline
\hline 
{\tiny{}BNC (RF - 50 Ohm) male/female} &  & {\tiny{}11/1} &  & {\tiny{}L-com} & {\tiny{}lead}\tabularnewline
\hline 
{\tiny{}wire solder } &  & {\tiny{}31.25/1} &  & {\tiny{}FCTA Mexico} & \tabularnewline
\hline 
{\tiny{}Acrylic materials} &  & {\tiny{}15/1 tube} &  & {\tiny{}AvanceyTec} & {\tiny{}acrylic}\tabularnewline
\hline 
{\tiny{}Ceramic trimmers ( 1.5 pF - 30 pF)} &  & {\tiny{}150} &  & {\tiny{}DigiKey} & {\tiny{}Polytetrafluoroethylene (PTFE)}\tabularnewline
\hline 
{\tiny{}Plastic C-clamps} &  & {\tiny{}4.51/1} &  & {\tiny{}Olson} & {\tiny{}plastic material}\tabularnewline
\hline 
 &  &  &  &  & \tabularnewline
\hline 
\end{tabular}
\par\end{center}

\section{Build instructions}
All in-house BC coil prototypes were constructed following a method similar to that described by Kemper et al. [15] for birdcage coil design. However, instead of utilizing a 3D printer, we opted for manual fabrication of the rungs and end rings by cutting them from copper sheets. These copper components were carefully shaped and then assembled onto acrylic cylinders to form the complete birdcage coil structure.
To ensure proper functionality, both non-magnetic chip capacitors and trimmers were soldered onto the coil to form high-pass and low-pass versions of the BC coil, depending on the specific design requirements.

The values of the passive components, including capacitors and trimmers, were initially estimated using the OpenBirdcageBuilder [16] software, a tool specifically developed to assist in the design and optimization of birdcage coils. This allowed us to obtain a preliminary set of values for the components needed to achieve the target resonant frequencies.
After component selection, fine-tuning and 50 $\Omega$ impedance matching were performed using a network analyzer to precisely adjust the coil's resonant frequency and impedance characteristics.

The process of tuning involved iterative adjustments, including the soldering and desoldering of various non-magnetic chip capacitors to achieve the desired resonant frequencies of 170 MHz, 300 MHz, and 650 MHz. This step was crucial for ensuring optimal performance of the coil in its intended application. The exact component values for each case, as well as the final tuning results, are provided in the preceding section of this document. Achieving accurate 50 $\Omega$ matching was particularly challenging and required careful adjustments to the passive components to ensure minimal reflection and maximum power transfer across the entire frequency range.

Throughout the fabrication and tuning process, careful attention was paid to maintaining the integrity of the coil\textquoteright s design and ensuring that the final prototype met the desired performance criteria. The iterative process of component selection, tuning, and impedance matching is essential for optimizing birdcage coils in applications such as MRI and RF spectroscopy, where precise resonance and impedance characteristics are critical for achieving high-quality results. 

The development process of a BC coil involves several stages: 

a) Calculation of the passive component values required to achieve the desired resonant frequency and 50 $\Omega$ impedance [16].
This step ensures that the coil operates efficiently at the intended frequency and matches the system's impedance. 

b) Construction of the BC prototype, which can be accomplished by either cutting the rungs and endrings from a copper sheet or utilizing a 3D printer. These fabrication techniques allow for precise coil geometries suited to the design specifications. 

c) and d) Iterative tuning and optimization: Once the prototype is assembled, the process of adding or removing components (such as capacitors or inductors) is used to fine-tune the coil to achieve the required 50 $\Omega$ impedance matching, ensuring minimal signal reflection and efficient energy transfer. 

e) Image acquisition: Once the coil is tuned, images are obtained to assess the coil's performance. 

f) Validation of the BC prototype: The final step involves validating the coil's performance by comparing experimental results with theoretical predictions (eq. 1), confirming that the coil meets the design goals and performs as expected in practical applications.

Fig. 5 illustrates a schematic diagram outlining the complete process of constructing the in-house BC coils and validating the theoretical magnetic field expression.

\medskip{}
\noindent\begin{minipage}[t]{1\columnwidth}
\begin{center}
\includegraphics[scale=1.25]{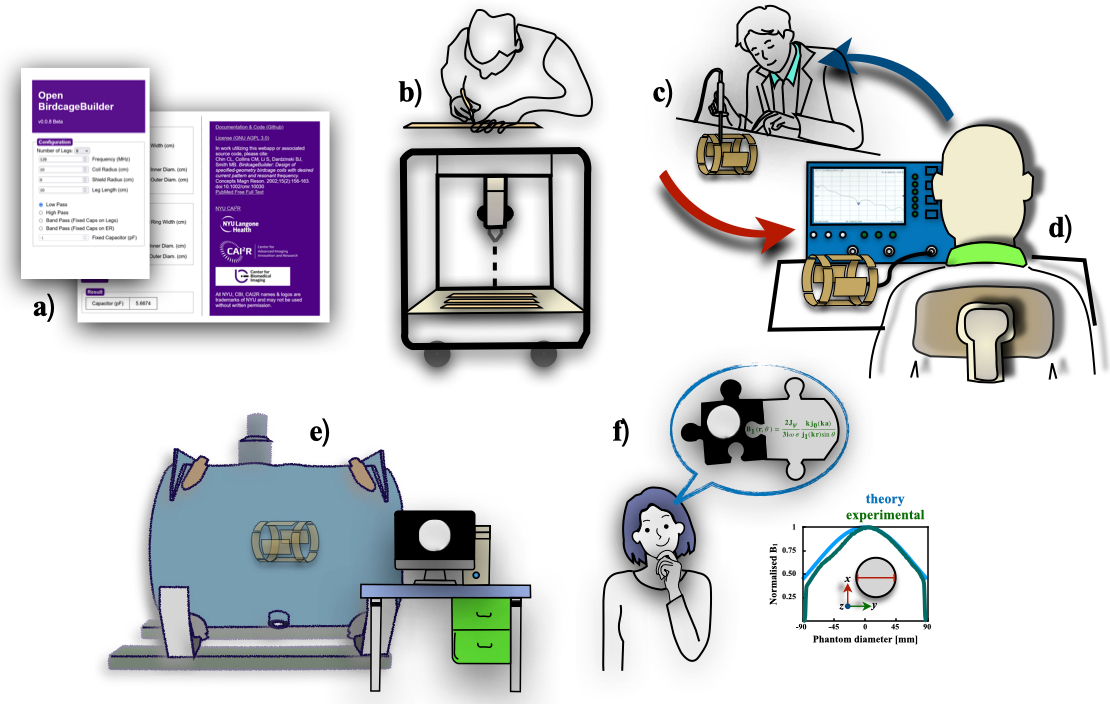}
\par\end{center}
\begin{center}
Fig. 2. Schematic of BC coil development process: a) Calculation of passive component values for resonant frequency and 50 $\Omega$ impedance [?]. b) Construction of the prototype using copper sheet or 3D printing. c) and d) Iterative tuning for 50 $\Omega$ impedance matching. e) Image acquisition. f) Validation through comparison of experimental and theoretical results.
\par\end{center}
\end{minipage}

\medskip{}

\section{Validation and characterization}
\subsection{Imaging experiments}

Phantom imaging experiments were conducted at various resonant frequencies using different acquisition parameters, pulse sequences, and MRI systems. 

15.2 T: Using the coil prototype shown in Fig. 1.c), imaging was performed with a saline solution-filled sphere (radius = 1 cm) utilizing a standard FLASH sequence. Acquisition parameters were as follows: TE/TR = 1.5 ms/120 ms, field of view (FOV) = 40 \texttimes{} 40 mm\texttwosuperior , matrix size = 256 \texttimes{} 256, slice thickness = 2 mm, and NEX = 1. A transceiver quadrature birdcage coil (Bruker Co., Ettlingen, Germany; model RES 650 1H 059/035 QSN TR, inner diameter = 35 mm, outer diameter = 59 mm, length = 30 cm) was used for these experiments.

7 T: Fig. 1.)a (In-house BC coil): A sphere (diameter = 2 cm) filled with a saline water solution was imaged on a research-dedicated 7T/21cm imager equipped with Direct Drive Technology (Agilent, Santa Clara, CA) using the in-house RF coil prototype. Phantom images were acquired using a standard gradient echo sequence with the following parameters: TR/TE = 1000/10 ms, FOV = 50 \texttimes{} 50 mm\texttwosuperior , matrix size = 256 \texttimes{} 256, slice thickness = 1 mm, NEX = 3, and flip angle = 20$^o$. Fig. 1.b) (In-house and vendor-provided BC coils): To validate the new coil design, imaging was performed using both the in-house BC coil and a vendor-provided birdcage coil (Bruker BioSpin MRI, GmbH, Ettlingen, Germany; model RF RES 300 1H 075/040 QSN TR, model no.: 1PT13161V3, serial no.: S0121, REV/VEC: 2P01.05).
Imaging experiments were conducted on a 7T/30cm Bruker imager using a standard gradient echo sequence. 

4 T: Using the coil prototype shown in Fig. 1.e), imaging was conducted in a whole-body superconducting magnet interfaced with an INOVA console (Varian, Inc., Palo Alto, CA, USA) and SONATA gradients (Siemens).
A spherical phantom (radius = 9 cm) was filled with distilled water containing a solution of creatine (50 mM), N-acetyl aspartate (12.5 mM), choline (3.0 mM), myo-inositol (7.5 mM), and glutamate (12.5 mM). Phantom images were acquired using the coil prototype and a spin-echo sequence with the following parameters: TR/TE = 3000/130 ms, FOV = 160 \texttimes{} 160 mm\texttwosuperior , matrix size = 256 \texttimes{} 256, slice thickness = 5 mm, and NEX = 1.

The summarized data from these experiments is presented in Table 1. 
\medskip{}

\newgeometry{left=2cm, right=2cm, top=0.5cm, bottom=0.5cm}

\begin{landscape}
\centering
\begin{adjustbox}{max width=1.3\textwidth}
\begin{tabular}{|c|p{4cm}|c|c|c|c|c|c|}
\hline 
\textbf{Reference} & \textbf{Acquisition parameters} & \textbf{$\mathrm{B}_{0}$} & \textbf{BC type/drive} & \textbf{Sequence} & \textbf{rungs/lenght/diameter [mm]} & \textbf{$Q_{l}/Q_{u}$ } & \textbf{$SNR/NF$} \tabularnewline
\hline 
\hline 
 & TR/TE = 1000/10 ms, FOV = 50$\mathrm{mm}^{2}$, matrix size
= 256x256, slice thickness = 1 mm, NEX = 3, flip angle = $20^{o}$ & 7 T & high-pass/quadrature & gradient echo & 16/105/50 & 35/39 &  \tabularnewline
\hline 
 & TE/TR = 25 ms/900 ms, FOV = 40 $\mathrm{mm}^{2}$, matrix
size = 256 \texttimes{} 256, slice thickness= 2 mm, NEX = 1 & 7 T & high-pass/quadrature & spin echo & 4/64/40 &  & /1.4 \tabularnewline
\hline 
 & TE/TR = 4.26/531.60 ms, FOV=100 $\mathrm{mm}^{2}$, matrix
size = 256 x 256, slice thickness = 1 mm, NEX = 10, & 7 T & transceiver mode/quadrature & spin echo & 6/80/50 & no record &  \tabularnewline
\hline 
 & TE/TR = 4.11 ms/530.70 ms, FOV = 800 $\mathrm{mm}^{2}$, matrix
size = 256 x 256, slice thickness = 2 mm, NEX = 3, flip angle = $20^{o}$, & 7 T & high-pass/quadarature & gradient echo & 4/113/66 & 43/105 & 76.54/2.6 \tabularnewline
\hline 
 & TE/TR=130/3000 ms, FOV=160 $\mathrm{mm}^{2}$, transversal
matrix size=256\texttimes 256, and saggital matrix size=128\texttimes 256,
slice thickness = 5 mm, NEX = 1 & 4 T & pass-band/quadrature & spin echo & 4/190/180 & 18.888/20.035 & \tabularnewline
\hline 
 & TE/TR = 1.5 ms/120 ms, FOV = 40 $\mathrm{mm}^{2}$, matrix
size = 256 x 256, slice thickness = 2 mm, NEX = 1, flip angle = $20^{o}$ & 15.2 T & transceiver mode/quadrature & gradient echo & 6/30/60 & no record & \tabularnewline
\hline 
\end{tabular}
\end{adjustbox}
\end{landscape}

\restoregeometry

Fig. 2 provides a photographs of the three MR imagers used to acquire the phantom images detailed above. These MRI systems were essential for evaluating the performance of the birdcage coil prototypes across different field strengths and imaging setups.

\medskip{}
\noindent\begin{minipage}[t]{1\columnwidth}
\begin{center}
\includegraphics[scale=2]{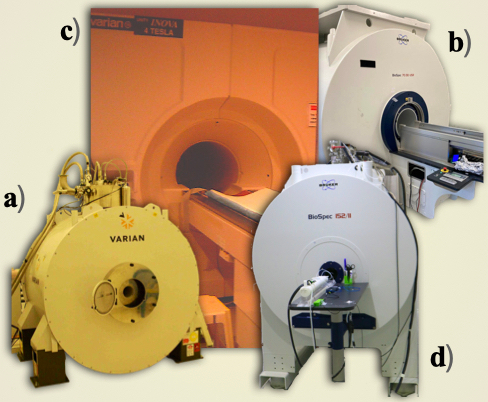}
\par\end{center}
\begin{center}
Figure 3. MRI systems used to experimentally test the in-house and vendor-provided birdcage (BC) coils at different resonant frequencies: a) 7 T imager (Agilent, Santa Clara, CA), b) 7 T imager and d) 15.2 T imager (Bruker BioSpin MRI, GmbH, Ettlingen, Germany), c) 4 T imager (Varian, Inc., Palo Alto, CA, USA and Siemens Healthineers AG, Erlangen, Germany).
\par\end{center}
\end{minipage}

Fig. 3 presents phantom images acquired using both in-house and vendor-provided BC coils in combination with different MR imagers. Overall, the images exhibit excellent quality, with clear and distinct representations of the phantom. However, some susceptibility artifacts are observed in Figs. 3.c), e), and f). These artifacts are typically associated with issues related to the construction and handling of the coil prototypes, which can lead to variations in the magnetic field distribution. Specifically, the artifact seen in Fig. 3.f) is caused by the proximity of a coaxial cable to the BC coil, which introduces local field distortions.

Despite these artifacts, it is important to emphasize that they did not have a significant impact on the overall image quality or the ability to interpret the data. The image resolution and clarity remained high across all cases, suggesting that the coil designs, while still prototypes, perform well even in the presence of minor imperfections.
Furthermore, the results demonstrate the robustness of the imaging system, as good image quality was maintained even when different BC coils were used with various MR imagers. This highlights the versatility and reliability of the approach, making it suitable for a wide range of applications, even in cases where coil construction may not be entirely optimized.

In summary, while susceptibility artifacts were present, their impact on the final image quality was minimal, and the overall performance of the BC coils across different MR systems was consistently strong, reinforcing the feasibility of using these coils in practical imaging scenarios.

\medskip{}
\noindent\begin{minipage}[t]{1\columnwidth}
\begin{center}
\includegraphics[scale=3]{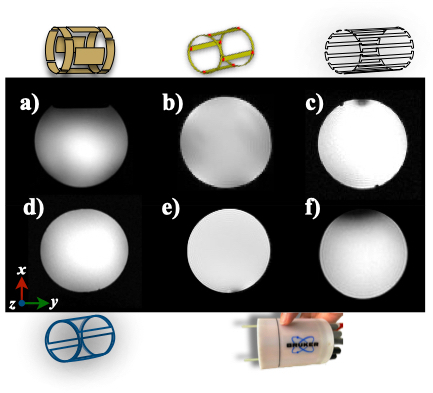}
\par\end{center}
\begin{center}
Fig 4. Phantom images acquired using the in-house and vendor-provided birdcage (BC) coils: a) 4 T, b)-e) 7 T, and f) 15.2 T, with the MRI systems described in Fig. 3.
\par\end{center}
\end{minipage}

\medskip{}

\subsection{Theoretical $\mathbf{B}_{1}$ validation}

To evaluate the performance of the various BC coils, image data were used to compute uniformity profiles, which were then compared to the theoretical $\mathrm{B}_{1}$ expression outlined in eq. (1). These uniformity profiles were measured along the red line indicated in Fig. 4.g), which served as the reference for consistent data acquisition across all cases. Using equ. (1), theoretical uniformity profiles were calculated for each BC coil, incorporating the coil's physical dimensions and its resonant frequency to model the expected magnetic field distribution.

\medskip{}
\noindent\begin{minipage}[t]{1\columnwidth}
\begin{center}
\includegraphics[scale=3]{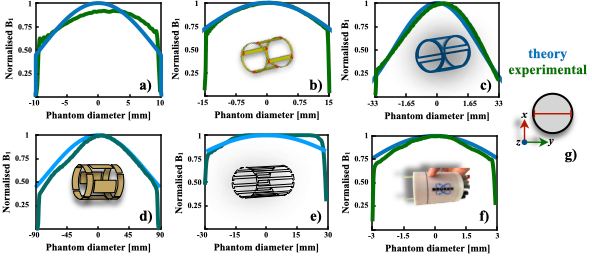}
\par\end{center}
\begin{center}
Fig. 4. Comparison of uniformity plots generated from the phantom image data in Fig. 3 and eq. (1) at magnetic field strengths of 4 T (d), 7 T (b, c, e, f), and 15.2 T (a). All profiles were acquired along the red line shown in (g).
\par\end{center}
\end{minipage}

\medskip{}

The comparison plots, shown in Fig. 4, illustrate the alignment between the experimental data and the theoretical predictions. A close match between the two indicates that the BC coils are performing as anticipated.
The experimental profiles derived from the imaging data align well with the theoretical profiles calculated using the formula in equ. (1), confirming that the coils are delivering the expected uniformity of the$\mathrm{B}_{1}$ field.

This strong agreement between theoretical and experimental results not only validates the performance of the coils but also supports the reliability of the theoretical model itself. Furthermore, the consistency observed in these results is in line with similar experimental and simulated findings reported in the literature [1,17-21]. This reinforces the accuracy of both the coil design and the theoretical framework used to predict coil behavior, providing confidence in the robustness and applicability of the BC coils for practical imaging applications. The results demonstrate that the coils maintain good uniformity across different setups, which is essential for achieving high-quality, reproducible imaging results.

This approach can serve as an alternative to burdensome numerical simulations of $\mathrm{B}_{1}$ using the Maxwell equations at ultra high frequencies. Coil dimensions can actually be computed previous to the construction of the coil prototype for optimal sensitivity, saving a great deal of time and effort. Additionally, commercial birdcage coils can be put to the test to investigate their sensitivities. This can particularly be of interest to those starting to build their own birdcage coil prototypes. The approach allows general design guidelines for the developments of birdcage coils for various preclinical applications at UHF MRI. 

This approach, in combination with cutting-edge software tools and 3D printing technologies, facilitates the design and fabrication of BC coils that are specifically optimized for various MRI applications.
By leveraging advanced simulation techniques and rapid prototyping, these coils can be precisely tailored to meet the unique needs of both preclinical and clinical MRI, improving image quality, spatial resolution, and overall diagnostic accuracy. This integrated methodology not only enhances the customization of coil geometries to accommodate different anatomical regions or specific clinical conditions but also enables more efficient production and testing, advancing the capabilities of MRI systems across a wide range of applications.

The experimental data were compared against the theoretical predictions, and the results demonstrated a strong correlation, confirming the reliability of the proposed model. This work provides a more accessible and robust framework for analyzing the magnetic field behavior in BC coils, which can be applied to the design and optimization of RF coils for a wide range of MRI applications.

\noindent
\textbf{CRediT author statement}\\
\noindent
{All authors contributed equally to the design, execution, analysis, and writing of this research. Each author played an integral role in the development of the study, and their combined efforts were essential to the successful completion of the work presented.}\\

\noindent
\textbf{Acknowledgements}\\
We sincerely appreciate the valuable comments and in-sights from the anonymous reviewers. We thank CONACYT Mexico (grant number 112092) and the Innovation and Research/Technology Support Program (PAPIIT) at UNAM (grant number IT 102116) for their support. Additionally, we are grateful for the assistance from the Preclinical Imaging Facility at the Vienna Biocenter Core Facilities and Dr. Scott Ireland from Bruker, Co.\\

\section{Appendix}

\subsection{Sphere model inside an uniform RF field}

According to London [?], we can consider a sphere with permeability $\mu$, permittivity $\varepsilon$, and electrical conductivity $\sigma$.
The sphere with $r=a$ is located at the origin of an uniform RF field given by $\mathbf{B}_{1}(t) = B_{1} \mathrm{e}^{-i\omega t} \hat{\mathbf{a}}_{z}$.
We can beging our analysis with the Maxwell-Ampere equation:

\[
\nabla\times\textbf{B}=\mu_{0}\mathbf{J}\quad(\mathrm{A}.1)
\]

where $\mathbf{J}$ represents the total current density in the system and $\mu_{0}$ is the permeability of vacuum. Eq. (A.1) describes how an electric current generates a magnetic field. In the context of spherical symmetry and under the influence of a radio frequency (RF) field applied to a dielectric material, we consider only the azimuthal component of the current density:

\[
\begin{array}{cc}
\textbf{J}_{\varphi}=f_{\phi}a_{\psi}=f(r)\sin\theta a_{\psi} & \quad(\mathrm{A}.2)\end{array}
\]

Here, $f(r)$ is the radial function that describes the spatial dependence of the current in the azimuthal direction within the sphere. It is defined by a second-order differential equation, which is subject to spherical symmetry. $f(r)$ satifies the following equation:

\[
\begin{array}{cc}
f''(r)+\frac{2}{r}f'(r)+\left(k^{2}-\frac{2}{r^{2}}\right)f(r)=0 & \quad(\mathrm{A}.3)\end{array}
\]

where $k=\sqrt{\mu\omega\left(\epsilon\omega+i\sigma\right)}$ is the propagation constant whose general solution is

\[
\begin{array}{cc}
f(r)=Aj_{1}(kr)+By_{1}(kr) & \quad(\mathrm{A}.4)\end{array}
\]

where$j_{n}(x)$ and $y_{n}(x)$ are the first-and-second-order spherical functions of Bessel. Now, $y_{1}(kr)$ diverges at $r=0$ so the coefficient $B=0$, this implies that:

Here, $j_{n}(x)$ and $y_{n}(x)$ are the first- and second-kind spherical Bessel functions, respectively. Since $y_{1}(kr)$ diverges at $r=0$, the coefficient $B=0$. This leads to the following conclusion:

\[
\begin{array}{cc}
f(r)=Aj_{1}(kr) & \quad(\mathrm{A}.5)\end{array}
\]

Consequently, the zimuathal current density is:

\[
\begin{array}{cc}
J_{\psi}=Aj_{1}(kr)\sin\theta & \quad(\mathrm{A}.6)\end{array}
\]

To obtain coefficient $A$ must satisfy the boundary condition at $r=1$ for the magnetic field within the sphere:

\[
\begin{array}{cc}
\mathrm{B}_{1r} & =\frac{2A}{i\omega\sigma r}j_{1}(kr)\cos\theta=\mathrm{B}_{1}+\frac{\mu_{0}m}{2\pi r^{3}}\quad(\mathrm{A}.7)\\
\mathrm{B}_{1\theta} & =\frac{-A}{i\omega\sigma r}(j_{1}(kr)+krj_{1}'(kr))\sin\theta=-\mathrm{B}_{1}+\frac{\mu_{0}m}{2\pi r^{3}}\quad(\mathrm{A}.8)
\end{array}
\]

and $k$ is the propagation constant and $j_{0}(kr)$ are the spherical Bessel functions. $\mathrm{B}_{1}(r)$ can be expressed in a more compact way, as follows [3]:

\[
\mathrm{B_{1}}=\frac{3\mathrm{B}_{\mathrm{center}}}{2}\left[j_{0}(kr)-\frac{j_{1}(kr)}{kr}\right]v+\left[3\frac{\mathit{j}_{1}(\mathit{kr})}{\mathit{kr}}-\mathit{j}_{0}(\mathit{kr})\right]\cdot\left(\mathbf{n}\cdot\mathbf{v}\right)\mathbf{n}\;(\mathrm{A}.9)
\]

A more detailed derivation of the sensitivity expression for a birdcage coil can be found in ref. [21,22]. $\mathbf{v}=\frac{1}{\sqrt{2}}\left(1,i,\mathrm{0}\right)$ is a unit vector in the direction of the circularly polarised RF field generated by the volume coil, $\mathbf{n}=\left(\sin\theta\cos\phi,\sin\theta\sin\phi,\mathrm{\cos\theta}\right)$ is the unit normal vector to the radial direction and $\mathbf{n}\cdot\mathbf{v}=\frac{\sin\theta}{\sqrt{2}}\left(\cos\phi+i\sin\phi\right)$.
The $B_{\mathrm{center}}$ is the magnetic field at the coil's isocenter relative to the end-ring current is [19,20]:

\[
\mathrm{B}_{\mathrm{center}}=\sin\left(\frac{\pi}{N}\right)\frac{Ni\left(l^{2}+2d^{2}\right)}{\pi d\left(l^{2}+d^{2}\right)^{3/2}}\qquad(\mathrm{A}.10)
\]

where \emph{l} and \emph{d} represent the length and the diameter of the volume coil, respectively, and Eq. (8) was computed with no shielding. Eq. (5) can be written in spherical coordinates, 

\[
\begin{array}{c}
\mathrm{B_{1,\mathit{x}}}=\left(\frac{3i\mathrm{B}_{\mathrm{center}}}{2\sqrt{2}}\left[j_{0}(kr)-\frac{j_{1}(kr)}{kr}\right]+\left[3\frac{\mathit{j}_{1}(\mathit{kr})}{\mathit{kr}}-\mathit{j}_{0}(\mathit{kr})\right]\sin^{2}\theta\cos^{2}\phi\right)\;(\mathrm{A}.11)\\
\\
\mathrm{B_{1,\mathit{y}}}=\left(\frac{3i\mathrm{B}_{\mathrm{center}}}{2\sqrt{2}}\left[j_{0}(kr)-\frac{j_{1}(kr)}{kr}\right]+\left[3\frac{\mathit{j}_{1}(\mathit{kr})}{\mathit{kr}}-\mathit{j}_{0}(\mathit{kr})\right]\sin^{2}\theta\sin^{2}\phi\right)\;(\mathrm{A}.12)\\
\\
\mathrm{B_{1,\mathit{z}}}=\left(\frac{3i\mathrm{B}_{\mathrm{center}}}{2\sqrt{2}}\left[3\frac{\mathit{j}_{1}(\mathit{kr})}{\mathit{kr}}-\mathit{j}_{0}(\mathit{kr})\right]\sin\theta\cos\phi\cos\theta\right)\;(\mathrm{A}.13)
\end{array}
\]

A more realistic scenario in MRI involves considering the presence of a dielectric material inside the sphere. The azimutahl current density inside the dielectric material can be written as

\[
J_{\psi}=\frac{3i\omega\sigma\mathrm{B}_{1}}{2}\frac{j_{1}(kr)}{kj_{0}(ka)}\sin\theta\;(\mathrm{A}.14)
\]

where $\omega$ is the resonant frequency, $\sigma$ is the conductivity, \emph{k} is the propagation constant, $\epsilon_{r}$ is the relative permitivitty, \emph{a} is the sphere radius. Finally, the magnetic field, $\mathrm{B}_{1}$ in spherical coordinates is

\[
\mathrm{B}_{1}\left(r,\theta\right)=\frac{2J_{\psi}}{3i\omega\sigma}\frac{kj_{0}(ka)}{j_{1}(kr)\sin\theta}\;(\mathrm{A}.1)
\]


\noindent 
\begin{thebibliography}{10}
\bibitem{key-29}Vaughan, J. T., \& Griffiths, J. R. (Eds.). (2012).
RF coils for MRI. John Wiley \& Sons.

\bibitem{key-28}Hayes CE. The development of the birdcage resonator:
a historical perspective. NMR Biomed. 2009;22(9):908-18. doi: 10.1002/nbm.1431. 

\bibitem{key-27}Ahmad SF, Kim YC, Choi IC, Kim HD. Recent Progress
in Birdcage RF Coil Technology for MRI System. Diagnostics (Basel).
2020, 27;10(12):1017. doi: 10.3390/diagnostics10121017. 

\bibitem{key-26}Seo JH, Han Y, Chung JY. A Comparative Study of Birdcage
RF Coil Configurations for Ultra-High Field Magnetic Resonance Imaging.
Sensors (Basel). 2022 Feb 23;22(5):1741. doi: 10.3390/s22051741. 

\bibitem{key-25}Choi CH, Bruch M, Hong SM, Krause S, Stegmayr C,
Schwan S, Worthoff WA, Felder J, Shah NJ. A Modified Quadrature Birdcage
Coil Incorporated With a Curved Feature for In Ovo MR Imaging. IEEE
Open J Eng Med Biol. 2024;28(5):534-541. doi: 10.1109/OJEMB.2024.342023.

\bibitem{key-24}Bidinosti CP, Chapple EM, Hayden, ME. The sphere
in a uniform RF field\textemdash Revisited. Concp Magn Reson Part
B: Magn Reson Eng. 2007;31(3):191-202. https://doi.org/10.1002/cmr.b.20090.

\bibitem{key-23}London F. 1961. Superfluids Volume 1: Macroscopic
Theory of Superconductivity, 2nd revised edition. New York: Dover;
First published New York: Wiley; 1950.

\bibitem{key-22}O. Marrufo, F. Vazquez, R. Martin, A. O. Rodriguez,
S. E. Solis-Najera. Double-crossed radiofrequency coil with improved
uniformity for rodent MRI at 7 T. Journal of Magnetic Resonance Open,
2022. 10.1016/j.jmro.2022.100068. 

\bibitem{key-21}S. Solis-Najera, R. Ruiz, R. Martin, F. Vazquez,
O. Marrufo, A. O. Rodriguez. A theoretical and experimental investigation
on a volume coil with slotted end-rings for rat MRI at 7 T. Magnetic
Resonance Materials in Physics, Biology and Medicine,  36, 911\textendash 919,
2023. https://doi.org/10.1007/s10334-023-01096-w. 

\bibitem{key-20}R. Martin, J. F. Vazquez, O. Marrufo, S. E. Solis,
A. Osorio, A. O. Rodriguez. SAR of a birdcage coil with variable number
of rungs at 300 MHz. Measurement 82, 482-489, 2016. http://dx.doi.org/10.1016/j.measurement.2016.01.013. 

\bibitem{key-19}Doty FD, Entzminger G, Kulkarni J, Pamarthy K, Staab
JP. Radio frequency coil technology for small-animal MRI. NMR Biomed.
2007 May;20(3):304-25. doi: 10.1002/nbm.1149. 

\bibitem{key-18}F. Vazquez, S. E. A. Villareal, J. Lazovic, R. Martin,
S. Solis-Najera, A. O. Rodriguez. RF coil that minimizes electronic
components while enhancing performance for rodent MRI at 7 Tesla.
10, 055040, 2024. Biomedical Physics and Engineering Express. https://doi.org/10.1088/2057-1976/ad7265. 

\bibitem{key-17}S. E. Solis, G. Cuellar, R. L. Wang, D. Tomasi, A.
O. Rodriguez, Transceiver 4-leg birdcage for high field MRI: knee
imaging. Revista Mexicana de Física, 52(3), 215-221, 2008. 

\bibitem{key-16}A. Villarreal, J. Zinnanti, S. Solis-Najera, F. Vazquez,
A. O. Rodriguez. Experimental validation of a B1 theoretical model
for a birdcage coil for preclinical UHF MRI. ISMRM Workshop on Ultra-High
Field MR, 2022. 

\bibitem{key-15}Kemper, P., Thoming, J., \& Kustermann, E. (2022).
Tailored birdcage resonator for magnetic resonance imaging at 7 T
using 3D printing. HardwareX, 12, e00326.

\bibitem{key-14}Chin CL, Collins CM, Li S, Dardzinski BJ, Smith MB.
BirdcageBuilder: Design of Specified-Geometry Birdcage Coils with
Desired Current Pattern and Resonant Frequency. Concepts Magn Reson.
2002 Jun;15(2):156-163. doi: 10.1002/cmr.10030.

\bibitem{key-13}Foo TK, Hayes CE, Kang YW. An analytical model for
the design of RF resonators for MR body imaging. Magn Reson Med. 1991;21(2):165-77.
doi: 10.1002/mrm.1910210202.

\bibitem{key-12}H. Vesselle and R. E. Collin, The signal-to-noise
ratio of nuclear magnetic resonance surface coils and application
to a lossy dielectric cylinder model. I. Theory. IEEE Trans Biomed
Eng.1995;42(5):497-506. doi: 10.1109/10.376154

\bibitem{key-11}Ocali O, Atalar E. Ultimate intrinsic signal-to-noise
ratio in MRI. Magn Reson Med. 1998;39(3):462-73. doi: 10.1002/mrm.1910390317. 

\bibitem{key-10}Riauka TA, De Zanche NF, Thompson R, Vermeulen FE,
Capjack CE, Allen PS. A numerical approach to non-circular birdcage
RF coil optimization: verification with a fourth-order coil. Magn
Reson Med. 1999 Jun;41(6):1180-8. doi: 10.1002/(sici)1522-2594(199906)41:6<1180::aid-mrm14>3.0.co;2-d. 

\bibitem{key-9}Hoult DI, Phil D. Sensitivity and power deposition
in a high-field imaging experiment. J Magn Reson Imaging. 2000;12(1):46-67.
doi: 10.1002/1522-2586(200007)12:1<46::aid-jmri6>3.0.co;2-d.

\bibitem{key-1}Celik H, Eryaman Y, Altinta\c{s} A, Abdel-Hafez IA,
Atalar E. Evaluation of internal MRI coils using ultimate intrinsic
SNR. Magn Reson Med. 2004;52(3):640-9. doi: 10.1002/mrm.20200. 

\bibitem{key-2}Ibrahim TS. Analytical approach to the MR signal.
Magn Reson Med. 2005;54(3):677-82. doi: 10.1002/mrm.20600.

\bibitem{key-3}Pfrommer A, Henning A. On the Contribution of Curl-Free
Current Patterns to the Ultimate Intrinsic Signal-to-Noise Ratio at
Ultra-High Field Strength. NMR Biomed. 2017;30(5). doi: 10.1002/nbm.3691. 

\bibitem{key-4}Cline H, Mallozzi R, Li Z, McKinnon G, Barber W. Radiofrequency
power deposition utilizing thermal imaging. Magn Reson Med. 2004;51(6):1129-37.
doi: 10.1002/mrm.20064. PMID: 15170832.

\bibitem{key-5}Jin, J., Shen, G., \& Perkins, T. (1994). On the field
inhomogeneity of a birdcage coil. Magnetic resonance in medicine,
32(3), 418-422.

\bibitem{key-6}Spence, D. K., \& Wright, S. M. (2003). 2-D full wave
solution for the analysis and design of birdcage coils. Concepts in
Magnetic Resonance Part B: Magnetic Resonance Engineering: An Educational
Journal, 18(1), 15-23.

\bibitem{key-7}Giovannetti, G., Landini, L., Santarelli, M. F., \&
Positano, V. (2002). A fast and accurate simulator for the design
of birdcage coils in MRI. Magnetic Resonance Materials in Physics,
Biology and Medicine, 15, 36-44. 

\bibitem{key-8}Nikulin, A., De Rosny, J., Haliot, K., Larrat, B.,
\& Ourir, A. (2019). Opencage radio frequency coil for magnetic resonance
imaging. Applied Physics Letters, 114(5).
\end{thebibliography}
\end{document}